# Exchange-dominated Standing Spin Wave Excitations under microwave irradiation in $Ni_{80}Fe_{20}$ Thin Films

Ziqian Wang, Xiaofeng Zhu, Xiaoshuang Chen, and Wei Lu

We investigated the microwave-assisted DC voltages of ferromagnetic resonances and exchange-dominated standing spin wave excitations in two different in-plane magnetized permalloy thin films via homodyne detection. The line shapes of ferromagnetic resonance spectra and the dispersion curves of ferromagnetic resonance and standing spin wave are in agreement of previous studies, while further investigations of DC voltage spectra for these two excitations reveal that 1. unlike ferromagnetic resonance signals, the anti-symmetrical line shapes of standing spin wave excitations are not depend on the electromagnetic relative phase of assisted microwave, and 2. linewidths of their DC voltage spectra are distinct. The complicated spin dynamics of standing spin wave is consequently discussed by applying Landau-Lifshitz-Gilbert equation in term of exchange interaction.

[1]National Laboratory of Infrared Physics, Shanghai Institute of Technical Physics, Chinese Academy of Sciences

500 Yutian Rd, Shanghai, 200083, China

DC effect in ferromagnetic conductors under microwave radiation has been firstly studied by *H. J. Juretschke* since 1960.[1-6] This solid-state phenomenon, generated by the galvanomagnetic effects and magnetization dynamics,[2,7] is usually electrically detected as non-zero time-averaged DC voltage by homodyne detection.[8-11] Homodyne detection of DC effect bears an important role for both scientific points of view and technological perspectives. Firstly, the signal spectrum of this effect is found to be very informative to further understand the spin dynamics: measured signal spectra have been applied to probe the microwave assisted magnetization switching,[12] which is considered as a pivotal factor for the next generation magnetic recording technology with faster speed.[13,14] In addition, researchers can obtain some important physical process, such as spin pumping, in ferromagnetic materials via this technology.[15-17] Furthermore, DC response of microwave reveals the electromagnetic features of assisted microwaves since both of the line shapes and amplitudes of signals are sensitive to the assisted microwave's frequency, polarization and electromagnetic relative phase (the phase difference between e-component and h-component of electromagnetic wave):[8,10] a novel microwave imaging technology is developed through analyzing the microwave induced DC signals.[18]

Most of recent studies and applications of electrically detected signals are based on the DC response of ferromagnetic resonance (FMR).[10,18,19] FMR is a typical excitation for uniform precession mode: every electron spin precesses at the same phase, frequency and amplitude. Spin waves, which exhibits as a non-uniform mode, is also possibly observed via microwave stimulation. However, fewer results have been reported for DC effect of spin wave due to the complicated spin-spin exchange and dipole interactions. Furthermore, the spin wave excitation is easily mixed with FMR. It is widely accepted that the interaction between spin and electromagnetic field is described by Landau-Lifshitz-Gilbert (LLG) equation, [20] the nonlinearlity of LLG equation indicates that simply extracting the spin wave signals from a whole DC signal spectrum by simply subtracting FMR component is not a property endeavor.

In ferromagnetic thin films, spin waves usually perform as standing spin waves (SSW) by taking boundary condition into account, [21-24] and, from a macroscopic view, result in magnetization precession. Although usually weaker than FMR signals, the excitation of spin waves are possible to be electrical detected since the time-dependent AMR is generated by precessional magnetization, the time-varying current is induced by microwave as well. To provide a general picture to show how the SSW mode evolves while it is excited, we are now focusing on the DC response of exchange-dominated SSW in ferromagnetic thin films. For overcoming the above predicament, two ferromagnetic films are fabricated as very thin microstrips for larger wave vector in order to avoid the coupling of SSW excitation and FMR, and the strip structures are designed for convenience of homodyne detection. The confusing line shapes of excited SSW spectra show that SSW mode is significantly different from FMR, not only in its non-uniformity of precession magnitudes. Similar phenomena occur in both of our two samples.

## Results

**Measurement setup, sample structures and magneto-resistance.** We show the preparation of our work in Figure 1. Fig. 1a is the schematic diagram of measurement, the coordinate system we select in our work is also shown in Fig. 1a and b. This work is performed on two $Ni_{80}Fe_{20}$ (permalloy, Py) thin films (labeled strip 1 and strip 2) that are illustrated in Fig. 1c, d. The dimensions of strip 1 are: length=2400μm, width=200μm and thickness=50nm

(prepared as described elsewhere) [8] while strip 2 is with a dimension of 300×7×0.1 μm$^3$ (prepared as described elsewhere). [9] These two strips are fabricated by lithography and liftoff technology on GaAs substrates. Unlike strip 1, strip 2 is buried by a coplanar waveguide (CPW) fabricated by a Cu/Cr (100nm) bilayer. A 200nm SiO$_2$ layer is also grown between CPW and strip 2 for electrical isolation. Two different methods are used to impose microwave for each strip: A rectangular waveguide is applied to transmit modulated microwave and ensure them normally propagate into strip 1, this rectangular waveguide is also applied as a holder for these two samples; for strip 2, localized microwave is generated by the CPW. The DC voltage signals of SSW and FMR are extracted by a lock-in amplifier connecting with two electrodes at both sides of each strip's length via gold bonding wires and coaxial cables. An electromagnet is employed to provide an external static magnetic field $\mu_o\mathbf{H_{ex}}$ in xz-plane with maximum amplitude of 1.5T, see in Fig. 1a. Signals are obtained on field-swept mode while the frequency of microwave is fixed when sweeping $\mathbf{H_{ex}}$. $\mathbf{H_{ex}}$ is in-plane with an angle $\theta$ to each strip's long axis. All data were obtained at room temperature.

The amplitude of FMR DC response is proportional to AMR ratio. [25-28] The AMR ratio for strip 1 (see in Fig. 1e) and strip 2 (see in Fig. 1f) are obtained by measuring the magnetoresistance $R$. AMR follows a simple relation of $\theta$ according to $R(\theta) = R_0 - \Delta R\sin^2\theta$ when the amplitude of $\mathbf{H_{ex}}$ is higher than a characteristic value to overcome the demagnetization field in x-direction and pull the magnetization $\mathbf{M}$ to $\theta$ direction. Fig. 1e, f reveal that the characteristic values are 20 Gauss for strip 1 and 140 Gauss for strip 2. The AMR ratio $\Delta R/(R_0-\Delta R)$ are 0.0167 for strip 1 and 0.0054 for strip 2.

**Distinct line shapes of FMR and SSW DC response.** Figure 2 shows the typical (a) FMR and (b) SSW DC responses ($V_{FMR}$ and $V_{SSW}$) in strip 1. The line shape of Fig. 2a is described as a sum of symmetrical component $L_{sym}=\Delta H_{FMR}/[(H_{ex}-H_{FMR})^2+\Delta H^2]$ and anti-symmetrical component $L_{a-sym}=(H_{ex}-H_{FMR})/[(H_{ex}-H_{FMR})^2+\Delta H_{FMR}^2]$, and is well fitted by $V_{FMR}\sim L_{a-sym}\times\sin\Psi+L_{sym}\times\cos\Psi$.() Here $H_{ex}=|\mathbf{H_{ex}}|$, $H_{FMR}$ is the resonant field of FMR, $\Delta H$ is the $V_{FMR}$ linewidth that related to $H_{ex}$ and damping, and $\Psi$ represents the relative phase of introduced microwave. Line shape of $V_{SSW}$ exhibits as anti-symmetry, as shown in Fig. 2b. The fitting curve in Fig.2b is obtained by applying $V_{SSW}\sim (H_{ex}-H_{SSW})/[(H_{ex}-H_{SSW})^2+\Delta H_{SSW}^2]$, where $H_{SSW}$ is the value of $H_{ex}$ while SSW is excited under certain microwave frequency and the phenomenological coefficient $\Delta H_{SSW}$ refers to the linewidth of $V_{SSW}$.

**FMR and SSW DC spectra.** Figure 3 shows the spectra of DC voltage $V_{DC}$ versus $H_{ex}$ imposed by microwave with different frequencies for (a) strip 1 and (b) strip 2 in field-swept mode. $V_{DC}$ is well fitted by applying the expression of $V_{DC}= V_{FMR}+V_{SSW}$. One SSW excitation and one FMR can be observed in each spectrum. All measured results in Figure 3 indicate that SSW line shapes are anti-symmetrical while FMR line shapes are combined by $L_{sym}$ and $L_{a-sym}$. The observed SSW signals in this article are 1$^{st}$ SSW. We does not observe higher ordered SSW excitations with larger wave vectors due to the confinement of equipment. Spectra of SSW excitation and FMR are split away in Fig. 3a, b. As noted in Figure 3, $H_{exchange}=H_{FMR}-H_{SSW}$ is represented as the magnitude of a static magnetic field $\mathbf{H_{exchange}}$, which refers to the exchange interaction between spins. $H_{exchange}$ is proportional to the wave vector of SSW. SSW in thicker Py sheet have larger wave vector, which results in a higher $H_{exchange}$ in strip 1 (1050Gauss) than that of strip 2 (300Gauss). Because of the smaller $H_{exchange}$, SSW excitations are slightly coupled with FMR in (b), while in (a), SSW and FMR curves are independent. The power of microwave induced to strip 1 (25dBm) on the output of an *rf* signal generator is higher than that to strip 2 (10dBm), also, AMR ratio in strip 1 is higher than that of strip 2. However spectra in strip 2 possess DC voltage amplitudes and better signal noise ratio (SNR), this is because CPW can introduce microwave to strip more efficiently than rectangular waveguide.

**Dispersion and line widths.** Figure 4 shows (a) the measured resonant $H_{ex}$ (including $H_{SSW}$ and $H_{FMR}$ for both strip 1 and 2), the comparison of $\Delta H_{SSW}$ and $\Delta H_{FMR}$ of (b) strip 1 and (c) strip 2 as functions of $f_{mw}$. Dispersion curve of FMR in (a) is well fitted by $2\pi f = \mu_0 \gamma \times [H_{ex} \times (H_{ex} + M/4\pi)]^{1/2}$ with magnetization moment $M = |\mathbf{M}|$, while SSW dispersions of strip 1 and 2 are fitted by $2\pi f = \mu_0 \gamma \times [(H_{ex} + H_{exchange}) \times (H_{ex} + H_{exchange} + M/4\pi)]^{1/2}$. In Fig. 2b, c we may observe that $\Delta H_{SSW} \neq \Delta H_{FMR}$.

## Discussion

Two factors that contribute to the DC response are represented in Fig. 5a: the microwave-induced time-varying current $i_{mw} = \varepsilon_{mw}\exp(i\omega t - i\Psi)/(R_0 - \Delta R)$ where $\varepsilon_{mw}$ is the electrical field of microwave that parallel to the strip's length and $\omega = 2\pi f$, and in-plane magnetized oscillating AMR $R(t) = R_0 - \Delta R + \Delta R \times \sin^2(\theta + m_x \exp(i\omega t - i\varphi)/M)$ where $m_x$ is the time-dependent *x*-component of precessional magnetization moment and $\varphi$ is the phase different between precession and microwave magnetic field $\mathbf{h_{mw}}$. By solving LLG equation:[20]

$$\frac{d\mathbf{M_n}}{dt} = \gamma\mu_0\left(\mathbf{H_{ex}} + \mathbf{h_d} + \mathbf{h_{mw}}\right) \times \mathbf{M_n} - \frac{\alpha}{M}\frac{d\mathbf{M_n}}{dt} \times \mathbf{M_n}, \tag{1}$$

$m_x$ and $\varphi$ are obtained. Here $\mathbf{h_d}$ is the time-varying demagnetization field along y-axis due to the oscillating y-component of $\mathbf{M}$, $\gamma$ is the gyromagnetic ratio and $\alpha$ is the Gilbert damping coefficient. In non-resonant uniformly precession mode, DC response is weak for the smaller precessional amplitude, as shown in Fig. 5b. The spin precession is strongly excited in FMR and results obviously voltage signal, see in Fig. 5c. FMR DC signal is resulted by $V_{FMR}= <i_{mw} \times R(t)>$, where $< >$ denotes the time averaging. In $\mathbf{H_{ex}}$-swept mode, the derived line shape of FMR is contributed by $L_{sym}$ and $L_{a-sym}$ by solving LLG equation, the measured FMR spectra in Figure 2, 3 and the dispersion in Fig. 4a verified this model. Similarly, $V_{SSW}$ is generated by $i_{mw}$ and $R(t)$ due to the coherent spin precession, as shown in Fig. 5d. Otherwise, if spin in SSW precesses incoherently, the DC response is weaker, see in Fig. 5e.

We begin the discussion on SSW with $\mathbf{H_{exchange}}$. In non-uniformly precession mode for a certain $\mathbf{M_n}$ in SSW, exchange interaction is induced by the nearest neighbor spins $\mathbf{M_{nb}} = \mathbf{M_{n-1}} + \mathbf{M_{n+1}}$. Consequently, LLG equation is transformed as:

$$\frac{d\mathbf{M_n}}{dt} = \gamma\mu_0\left(\mathbf{H_{ex}} + \mathbf{h_d} + \mathbf{h_{mw}}\right)\times\mathbf{M_n} - \frac{\alpha}{M}\frac{d\mathbf{M_n}}{dt}\times\mathbf{M_n} + \gamma\mu_0\lambda\mathbf{M_{nb}}\times\mathbf{M_n} \qquad (2)$$

here $\lambda$ is the Weiss molecular field constant. The final term in the right side of Eq. 2 expresses the exchange interaction on $\mathbf{M_n}=(m_{x\_n}\exp(i\omega t-i\varphi), m_{y\_n}\exp(i\omega t-i\varphi+i\phi), M)$ from $\mathbf{M_{nb}}=(m_{x\_nb}\exp(i\omega t-i\varphi), m_{y\_nb}\exp(i\omega t-i\varphi+i\phi), M)$, here $\phi$ represents the phase between $x$- and $y$-components of precessional $\mathbf{M}$. We are not going to discuss $\phi$ in this article since in the in-plane magnetization configuration, DC response not determined by $m_y$. Focusing on the exchange term of Eq. 2, the influence of $\mathbf{M_{nb}}$ acts on $\mathbf{M_n}$ is equivalent to a static magnetic field with its direction to $z$-axis. Especially, if each precessional motion in SSW follows sinusoidal distribution, the static field are the same for each spin and it is equal to $\mathbf{H_{exchange}}=(0, 0, 2\lambda M(1-\cos(\pi/N)))$. Noting that $N$ is the number of spin within half wavelength of SSW, as shown in Fig. 1a. Therefore, the impact of exchange interaction on SSW mode can be treat as adding $\mathbf{H_{exchange}}$ into $\mathbf{H_{ex}}$ for spins, if the assumption "each spin precesses at the same $\varphi$ and different $m_x$ in SSW" is true. Seemingly, the dispersion curves in Fig. 4a confirms this assumption. However, it is difficult by using the "symmetrical and anti-symmetrical" model to analyze the line shapes of $V_{SSW}$ since the expected symmetrical component is not observable. The resultant relative electromagnetic phase by applying the above model on SSW signal is $\Psi = \pm 90°$. This conclusion is misleading: $\Psi$ is fixed in field swept mode while the calculated $\Psi$ are not at $\pm 90°$ form $V_{FMR}$ analysis. As a result, shapes of $V_{SSW}$ spectra are not determined by $\Psi$.

Noting that what are also indicated by solving LLG equation: for two spins share the same $\alpha$, time-varying fields **h** (e.g. **h**$_{mw}$) and time-independent fields **H**$_{dc}$ (such as **H**$_{ex}$ and **H**$_{exchange}$), their $\varphi$ and $m_x$ will be equal; in contrast, once **H**$_{dc}$ are different while **h** are the same, their $\varphi$ will be different. These two derivations from LLG equation are described in the diagram of $m_x$ and $\varphi$ as a function of **H**$_{dc}$ in Fig. 6a. Thereby, "precession at same $\varphi$" indicates the same $m_x$ for each spin under same **H**$_{dc}$ and **h**. Thus, at least in resonant condition, this standing wave mode cannot exist while precession is enhanced by sweeping $H_{ex}$ into the vicinity of $H_{SSW}$. The increased non-uniformity of $m_x$ in SSW indicates the non-uniformity of $\varphi$, as illustrated in the resonant region of Fig. 6a. Theoretically, if each spin coherently precesses, the exchange interaction can be mathematically equivalent to **H**$_{exchange}$ without time-dependent item. In the ideal condition that $m_x$ distribution is sinusoidal, it is impossible for each spin precesses with different $m_x$ since their **H**$_{dc}$ and **h** are the same. Even though, if $m_x$ distribution in SSW is not sinusoidal, $H_{dc}=|$ **H**$_{dc}$ $|$ for each spin is distinct, while still, contains only time-independent item. As what has been derived from LLG equation, precessional phases for spins are still impossible to be uniform. Experimentally, if SSW always performs as standing wave mode, whatever in resonant or non-resonant condition, the line shapes of SSW DC response is expected to be the same as FMR: a combination of symmetric and anti-symmetric components. Nonetheless, the resultant data goes against this expectation since the shapes of SSW spectra are anti-symmetric. Two possible spintronic states of resonant SSW are: 1. SSW exhibits as a state with non-uniform $m_x$ and non-uniform $\varphi$ for each spin, as shown in Fig. 5e and 2. SSW degenerates to a non-resonant uniform precession state with weak $m_x$, as shown in Fig. 5c. The anti-symmetric SSW spectra are in good agreement of this assumption: $V_{SSW}=0$ at $H_{ex}=H_{SSW}$ while for these two possible states, there DC response are weak.

In contrast, non-resonant SSW performs as a standing wave because even weak $m_x$ for every spin might be distinct with each other, precessional motion in SSW is coherent since in the non-resonant condition, the value of $\varphi$ is approximately 0 for $H_{dc}>H_{FMR}$ and -180° for $H_{dc}<H_{FMR}$ where $H_{dc}=|\mathbf{H_{dc}}|$. In the non-resonant region, $V_{SSW}$ should be contributed by both symmetric and anti-symmetric components while the shape line is dominated by anti-symmetric component in non-resonant condition, as shown in Fig. 6b. That is why the shapes of SSW DC spectra are anti-symmetrical. In the resonant region, SSW cannot maintain its standing wave state, which results in the decrement of $V_{SSW}$ when $H_{ex}$ is approaching to $H_{SSW}$; while in the non-resonant state, microwave-assisted SSW is a standing wave with similar line shape of DC response that be dominated by anti-symmetric component. It is also implied that, although SSW DC spectra are anti-symmetric and they can be perfectly fitted by $V_{SSW} \sim (H_{ex}-H_{SSW})/[(H_{ex}-H_{SSW})^2+\Delta H_{SSW}^2]$, this formula is still empirical without the same physical meaning of $L_{a-sym}$. Furthermore, because of the influence of exchange interaction, the linewidth $\Delta H_{SSW}$ different from $\Delta H_{FMR}$ at the same $f_{mw}$, as shown in Fig. 4b, c.

In summary, spin wave mode and uniform precession mode are different, especially when they are excited by microwave. Non-resonant standing spin wave is indeed in the spin wave state, while SSW cannot exhibits as a spin wave in resonant condition because of the requirement of LLG equation. In contrast, each spin in uniform precession mode shares the same precessional amplitude and phase, whatever in the resonant (FMR) or non-resonant state. These are why the observed SSW spectra are anti-symmetrical and distinct from FMR spectra. Although SSW signals can be fitted very well by $L_{a\text{-}sym}$ that is used for analyzing FMR signal, the physical meaning of the formula that applied to fit $V_{SSW}$ is different from $V_{FMR}$, which finally leads to the difference between the linewidths of SSW and FMR.

# References

1    Juretschke, H. J. ELECTROMAGNETIC THEORY OF DC EFFECTS IN FERROMAGNETIC RESONANCE. *Journal of Applied Physics* **31**, 1401-1406, doi:10.1063/1.1735851 (1960).
2    Juretschke, H. J. DC DETECTION OF SPIN RESONANCE IN THIN METALLIC FILMS. *Journal of Applied Physics* **34**, 1223-&, doi:10.1063/1.1729445 (1963).
3    Egan, W. G. & Juretschke, H. J. DC DETECTION OF FERROMAGNETIC RESONANCE IN THIN NICKEL FILMS. *Journal of Applied Physics* **34**, 1477-&, doi:10.1063/1.1729604 (1963).
4    Moller, W. M. & Juretsch.Hj. DETERMINATION OF SPIN-WAVE BOUNDARY CONDITIONS BY DC EFFECTS IN SPIN-WAVE RESONANCE. *Physical Review B* **2**, 2651-&, doi:10.1103/PhysRevB.2.2651 (1970).
5    Costache, M. V., Watts, S. M., Sladkov, M., van der Wal, C. H. & van Wees, B. J. Large cone angle magnetization precession of an individual nanopatterned ferromagnet with dc electrical detection. *Applied Physics Letters* **89**, doi:10.1063/1.2400058 (2006).
6    Mizukami, S., Nagashima, S., Yakata, S., Ando, Y. & Miyazaki, T. Enhancement of DC voltage generated in ferromagnetic resonance for magnetic thin film. *Journal of Magnetism and Magnetic Materials* **310**, 2248-2249, doi:10.1016/j.jmmm.2006.10.827 (2007).
7    Weiss, H. GALVANOMAGNETIC DEVICES. *Ieee Spectrum* **5**, 75-& (1968).
8    Mecking, N., Gui, Y. S. & Hu, C. M. Microwave photovoltage and photoresistance effects in ferromagnetic microstrips. *Physical Review B* **76**, doi:10.1103/PhysRevB.76.224430 (2007).
9    Zhu, X. F. *et al.* Nonresonant spin rectification in the absence of an external applied magnetic field. *Physical Review B* **83**, doi:10.1103/PhysRevB.83.140402 (2011).
10   Zhu, X. F. *et al.* Dielectric measurements via a phase-resolved spintronic technique. *Physical Review B* **83**, doi:10.1103/PhysRevB.83.104407 (2011).
11   Chen, L., Matsukura, F. & Ohno, H. Direct-current voltages in (Ga,Mn)As structures induced by ferromagnetic resonance. *Nature Communications* **4**, doi:10.1038/ncomms3055 (2013).
12   Fan, X. *et al.* Electrical detection of microwave assisted magnetization switching in a Permalloy microstrip. *Applied Physics Letters* **95**, doi:10.1063/1.3200239 (2009).
13   Thirion, C., Wernsdorfer, W. & Mailly, D. Switching of magnetization by nonlinear resonance studied in single nanoparticles. *Nature Materials* **2**, 524-527, doi:10.1038/nmat946 (2003).
14   Katine, J. A., Albert, F. J., Buhrman, R. A., Myers, E. B. & Ralph, D. C. Current-driven magnetization reversal and spin-wave excitations in Co/Cu/Co pillars. *Physical Review Letters* **84**, 3149-3152, doi:10.1103/PhysRevLett.84.3149 (2000).
15   Costache, M. V., Sladkov, M., Watts, S. M., van der Wal, C. H. & van Wees, B. J. Electrical detection of spin pumping due to the precessing magnetization of a single ferromagnet. *Physical Review Letters* **97**, doi:10.1103/PhysRevLett.97.216603 (2006).
16   Costache, M. V., Watts, S. M., van der Wal, C. H. & van Wees, B. J. Electrical detection of spin pumping: dc voltage generated by ferromagnetic resonance at ferromagnet/nonmagnet contact. *Physical Review B* **78**, doi:10.1103/PhysRevB.78.064423 (2008).

17	Azevedo, A., Leao, L. H. V., Rodriguez-Suarez, R. L., Oliveira, A. B. & Rezende, S. M. Dc effect in ferromagnetic resonance: Evidence of the spin-pumping effect? *Journal of Applied Physics* **97**, doi:10.1063/1.1855251 (2005).
18	Bai, L. H. *et al.* The rf magnetic-field vector detector based on the spin rectification effect. *Applied Physics Letters* **92**, doi:10.1063/1.2837180 (2008).
19	Kittel, C. ON THE THEORY OF FERROMAGNETIC RESONANCE ABSORPTION. *Physical Review* **73**, 155-161, doi:10.1103/PhysRev.73.155 (1948).
20	Gilbert, T. L. A phenomenological theory of damping in ferromagnetic materials. *Ieee Transactions on Magnetics* **40**, 3443-3449, doi:10.1109/tmag.2004.836740 (2004).
21	Herring, C. & Kittel, C. ON THE THEORY OF SPIN WAVES IN FERROMAGNETIC MEDIA. *Physical Review* **81**, 869-880, doi:10.1103/PhysRev.81.869 (1951).
22	Mruczkiewicz, M. *et al.* Standing spin waves in magnonic crystals. *Journal of Applied Physics* **113**, doi:10.1063/1.4793085 (2013).
23	Cao, Y., Urazuka, Y., Tanaka, T. & Matsuyama, K. Quantized Standing Spin Waves in a Fabry-Perot Type Resonator Consisting of Magnetic Nanowires. *Ieee Transactions on Magnetics* **48**, 4317-4319, doi:10.1109/tmag.2012.2200241 (2012).
24	Jacobi, D. M. *et al.* Angular and frequency dependence of standing spin waves in FePt films. *Journal of Applied Physics* **111**, doi:10.1063/1.3682104 (2012).
25	Azevedo, A., Vilela-Leao, L. H., Rodriguez-Suarez, R. L., Santos, A. F. L. & Rezende, S. M. Spin pumping and anisotropic magnetoresistance voltages in magnetic bilayers: Theory and experiment. *Physical Review B* **83**, doi:10.1103/PhysRevB.83.144402 (2011).
26	Liu, Y. F., Cai, J. W. & Sun, L. Large enhancement of anisotropic magnetoresistance and thermal stability in Ta/NiFe/Ta trilayers with interfacial Pt addition. *Applied Physics Letters* **96**, doi:10.1063/1.3334720 (2010).
27	Tripathy, D., Vavassori, P., Porro, J. M., Adeyeye, A. O. & Singh, N. Magnetization reversal and anisotropic magnetoresistance behavior in bicomponent antidot nanostructures. *Applied Physics Letters* **97**, doi:10.1063/1.3474802 (2010).
28	Wang, S. *et al.* Anisotropic magnetoresistance of Ni81Fe19 films on NiFeNb buffer layer. *Journal of Alloys and Compounds* **575**, 419-422, doi:10.1016/j.jallcom.2013.05.200 (2013).

# Figure Captions

**Figure 1 | measurement setup, sample structures and magnetoresistance** (**a**) Schematic diagram of the experimental setup, a coordinate system selected for further calculation is also illustrated: x-axis lies parallel to **H**$_{ex}$ while y-axis is set as normal to strip's plane. The microwave-induced time-varying current $i_{mw}$ flows along strip's length. (**b**) Description of precessional *N* electron spins in a SSW. They are marked from **M**$_1$, to **M**$_N$. (**c**) Sample structure of a single Py thin strip (marked as strip 1). Modulated radio-frequency (*rf*) signal is introduced through a rectangular waveguide. (**d**) Structure of the sample with a Py thin strip (strip 2) that underneath a co-planar waveguide. Magnetoresistance measurements at orientations $\theta$=0 and 90° for (**e**) strip 1 ($R_0$=89.22 Ω, $\Delta R$=1.49 Ω) and (**f**) strip 2 ($R_0$=121.53Ω, $\Delta R$=0.66 Ω).

**Figure 2 | distinct line shapes of FMR and SSW DC responses.** The experimental (dots) and fitted (solid lines) signal of (**a**) a typical FMR DC response versus external field and (**b**) a characteristic anti-symmetric DC voltage signal in the vicinity of SSW excitation obtained at magnetic field angle $\theta$=45° and *rf* signal frequency *f*=11.7GHz with a modulation frequency $f_m$=5.27kHz in strip 1. The colored regions in panel **a** show the fitting of symmetrical (blue) and anti-symmetrical (red) components that contribute to the line shape of FMR spectrum.

**Figure 3 | FMR and SSW DC voltage spectra.** The experimental (grey dots) and fitted (colored lines) of (**a**) SSW and FMR DC voltage spectra of strip 1 obtained at microwave power P=25dBm and frequencies ranging from 12.3GHz to 17.2GHz in field-

swept mode and (**b**) SSW and FMR spectra of strip 2 obtained at microwave power P=10dBm and frequencies ranging from 6.5GHz to 9.5GHz in field-swept mode. $\theta$ is always fixed at 45°. The exchange interaction built-in field $H_{exchange}$ is also shown in panel **a** and **b**.

**Figure 4 | dispersion and linewidths.** (**a**) Measured dispersion (solid squares for strip 1 and hollow squares for strip 2) of FMR (orange squares) and SSW (green squares) of these two strips with their fitting results (solid lines). The linewidth $\Delta H$ for FMR (orange) and SSW (green) of (**b**) strip 1 and (**c**) strip 2 at different microwave frequencies.

**Figure 5 | diagrams of discussions.** (**a**) The illustration shows how $V_{DC}$ generates and 4 states that might caught $V_{DC}$: (**b**) resonant uniformly precession mode (FMR), (**c**) non-resonant uniformly precession mode, (**d**) standing wave mode and (**e**) non-uniform precession mode with the existence of phase difference within spins.

**Figure 6 | diagrams of discussions.** (**a**) Theoretical $\varphi$ and $m_x$ diagram for a precessional spin excited by static and time-varying magnetic field. (**b**) Theoretical line shape of FMR DC response. Noting that in the non-resonant region of **b**, the line shape is dominated by the anti-symmetric component while in the resonant region, the voltage is contributed by both $L_{sym}$ and $L_{a-sym}$.

Figure 1.

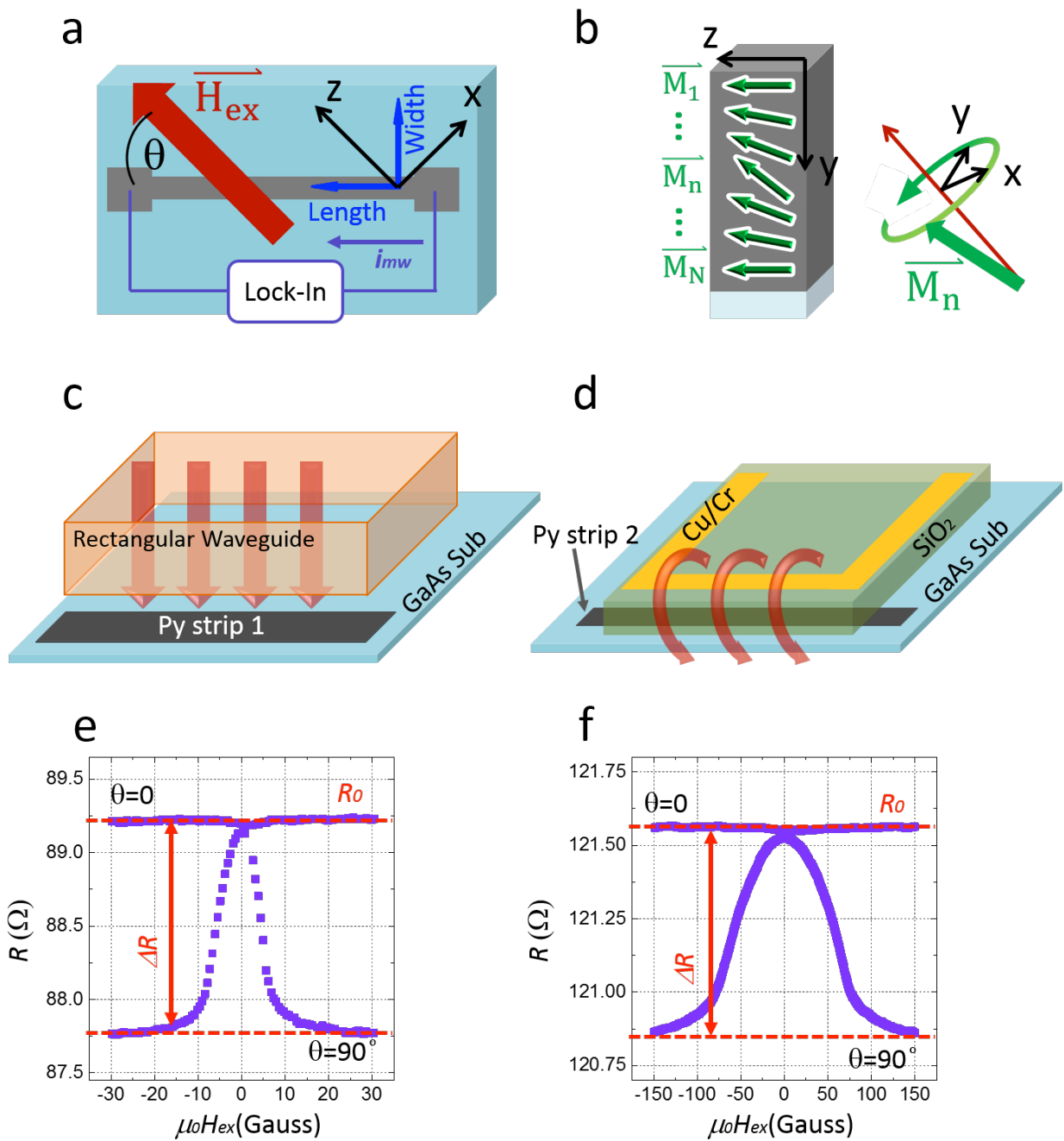

Figure 2.

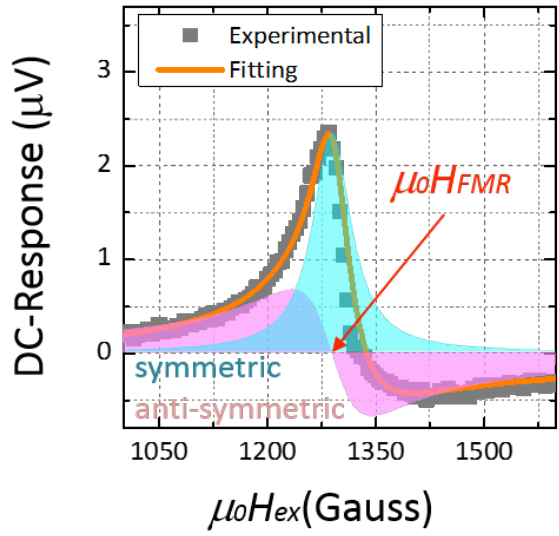 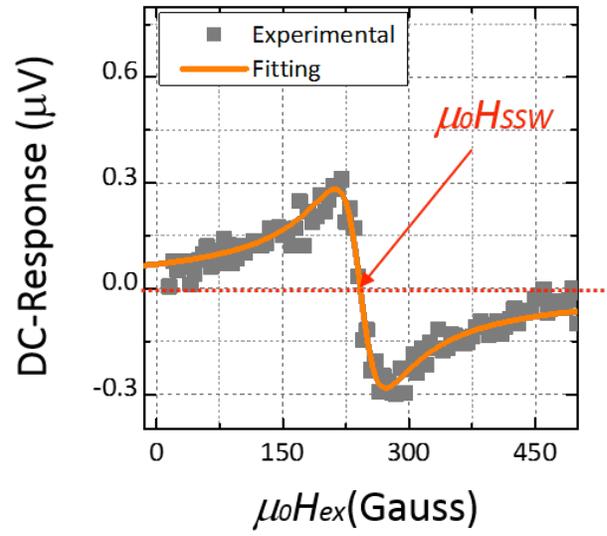

Figure 3. (double column)

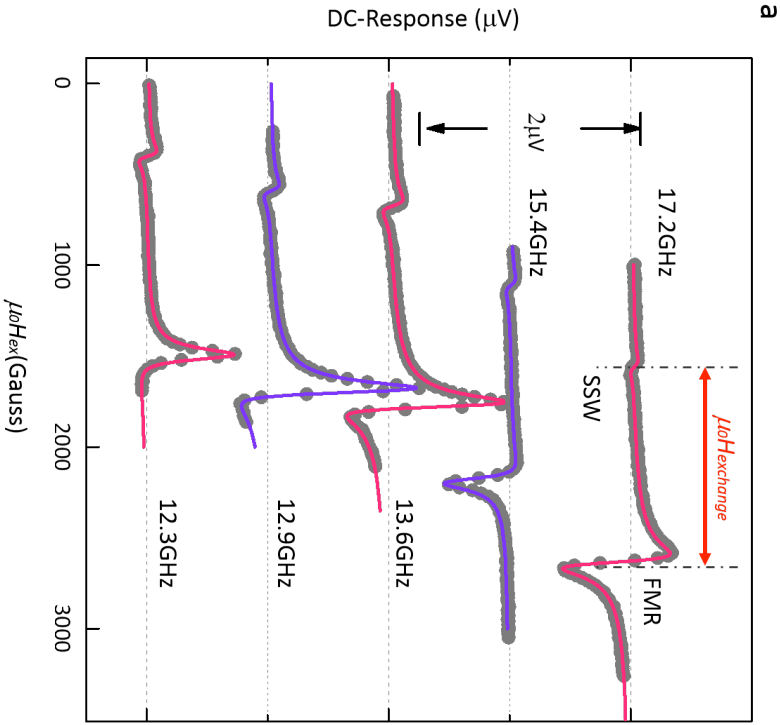
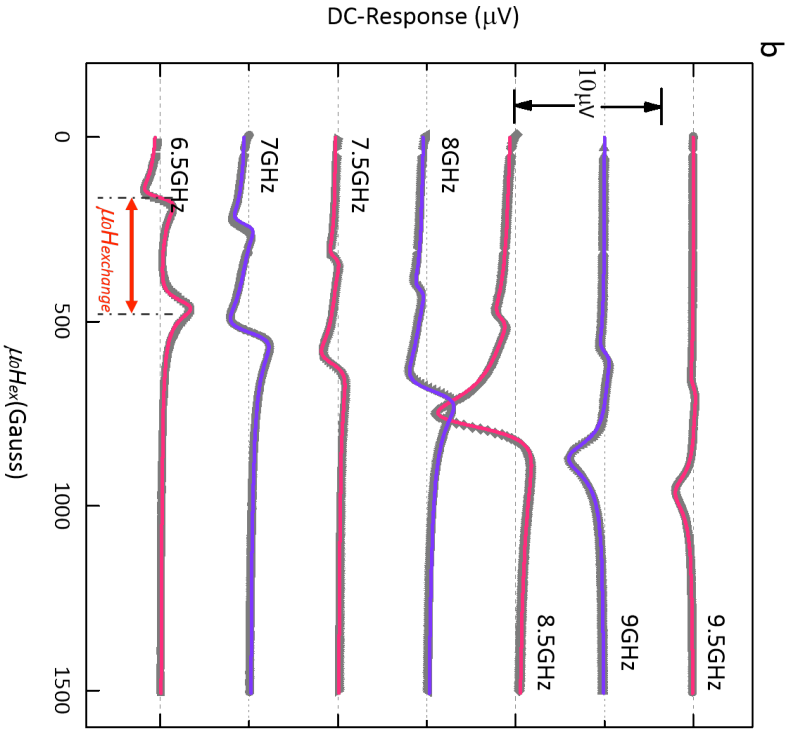

Figure 4.

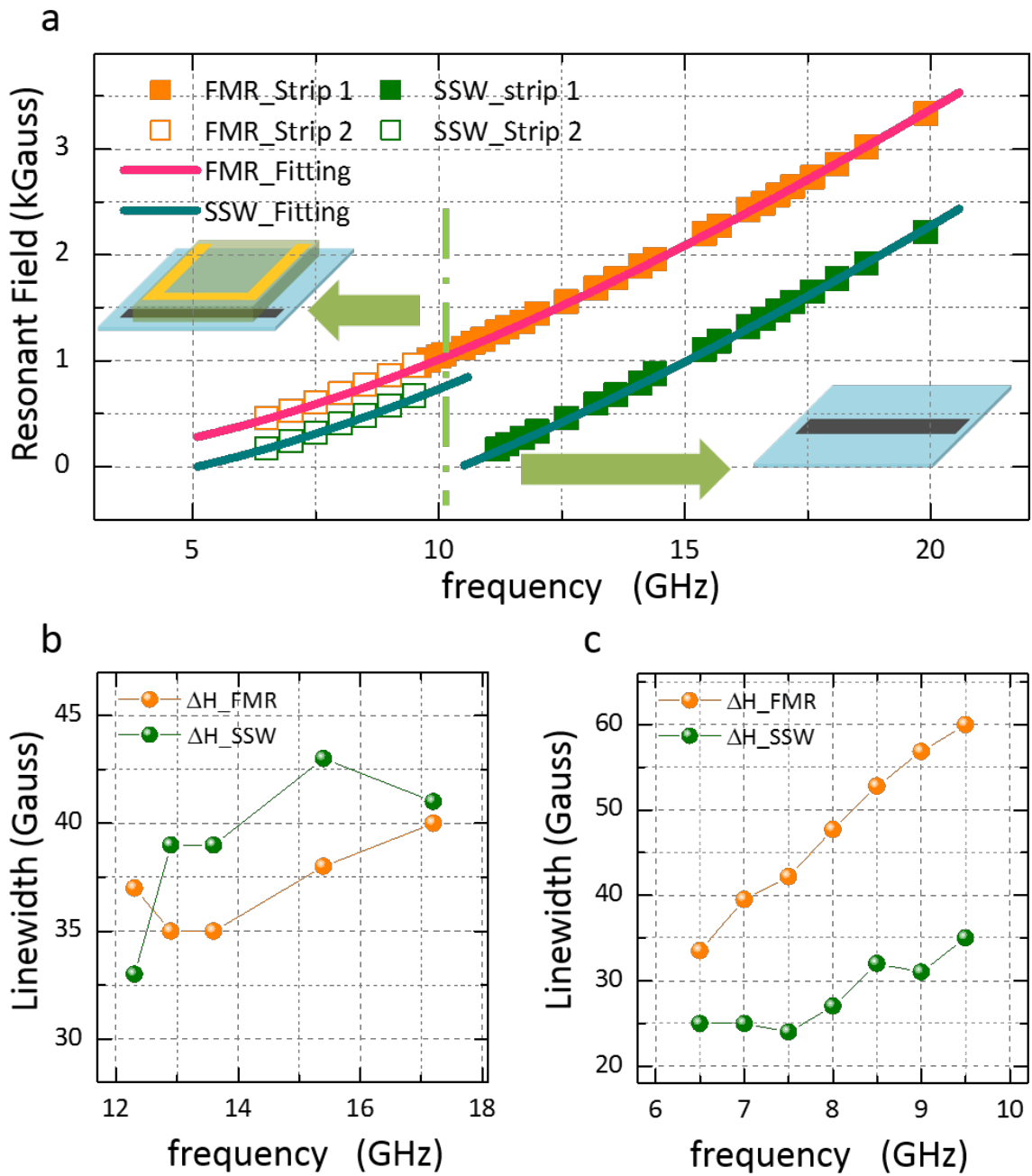

Figure 5.

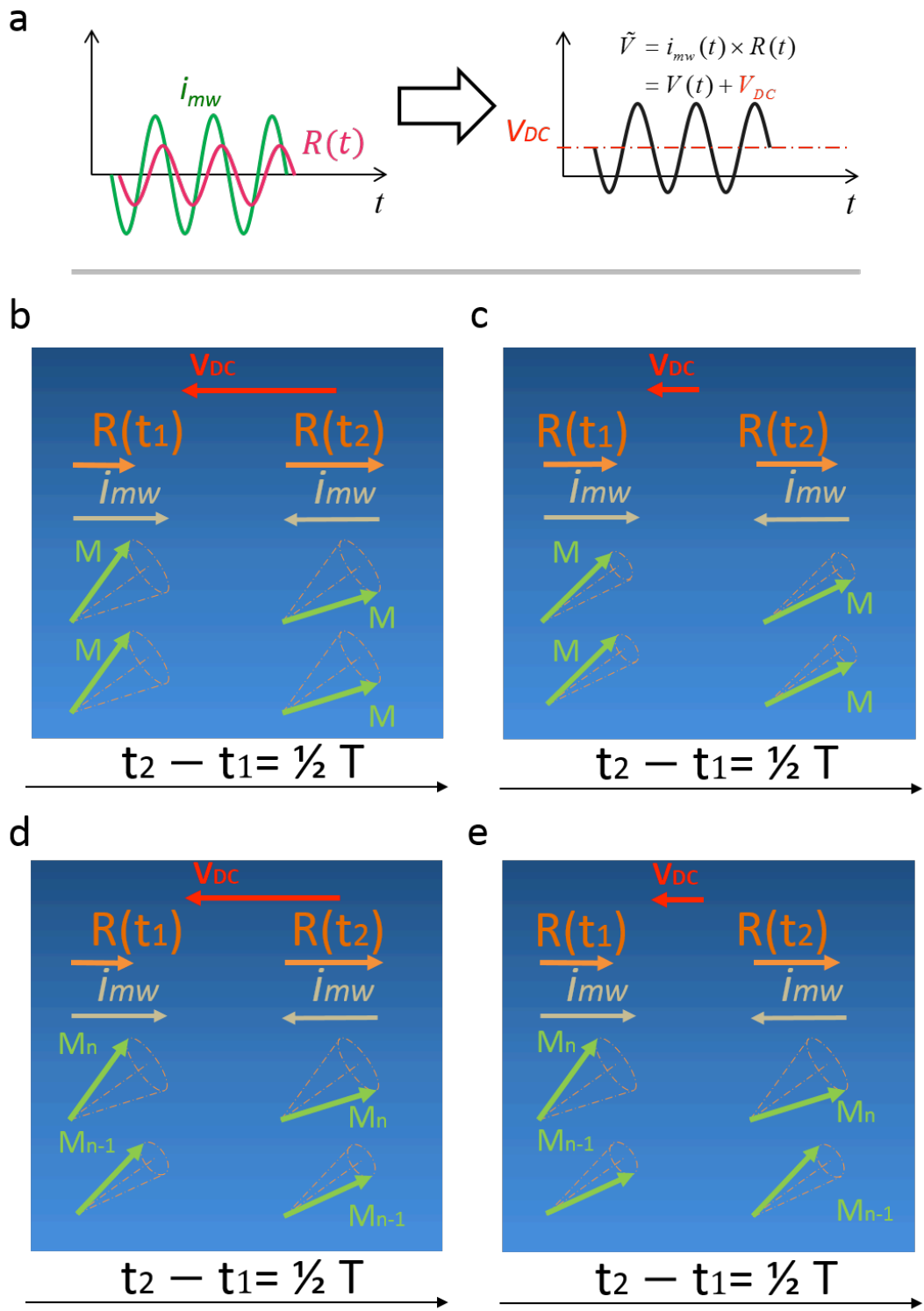

Figure 6.

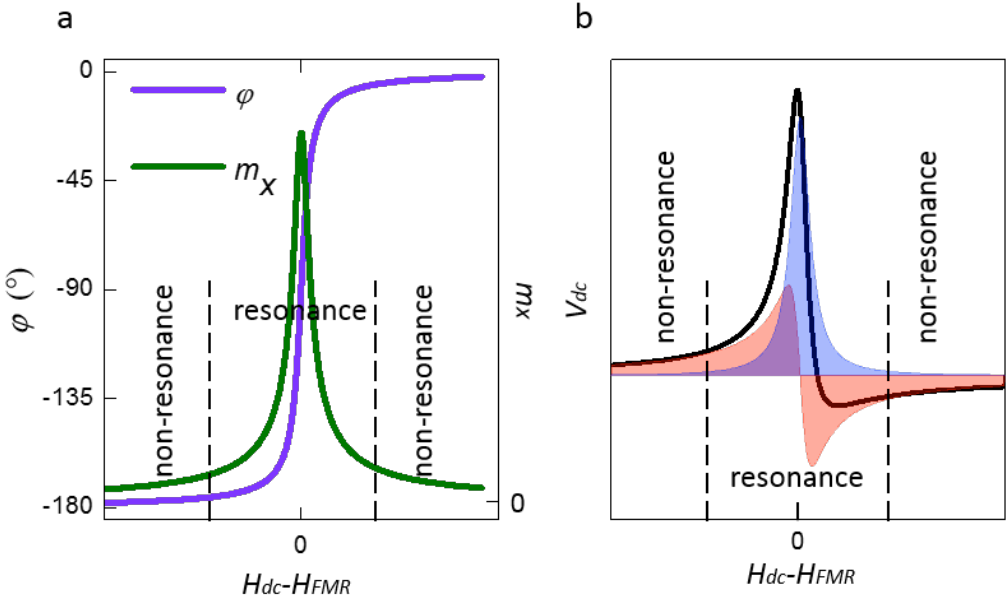